\documentclass[aps,prd,twocolumn,superscriptaddress,nofootinbib,showpacs,preprintnumbers]{revtex4}
\usepackage[english]{babel}

\begin{document}
\preprint{IFF-RCA-09-03}
\title{Lorentzian wormholes generalizes thermodynamics still further.}

\author{Prado Mart\'{\i}n-Moruno}
\email{pra@imaff.cfmac.csic.es}
\affiliation{Colina de los
Chopos, Instituto de F\'{\i}sica Fundamental, \\
Consejo Superior de Investigaciones Cient\'{\i}ficas, Serrano 121,
28006 Madrid, Spain}
\author{Pedro F. Gonz\'{a}lez-D\'{\i}az}
\email{p.gonzalezdiaz@imaff.cfmac.csic.es }
\affiliation{Colina de los
Chopos, Instituto de F\'{\i}sica Fundamental, \\
Consejo Superior de Investigaciones Cient\'{\i}ficas, Serrano 121,
28006 Madrid, Spain}

\begin{abstract}
This paper deals with some thermodynamical aspects of Lorentzian
wormholes, including the formulation of the three main laws and
the consideration of a possible thermal emission made up of some
sort of phantom radiation coming out from the wormhole at a
negative temperature. In order for these topics to be consistently
developed we have used a 2+2 formalism first advanced by Hayward
for spherically symmetric space-times, where a generalized surface
gravity is defined on the trapping horizon. Our results generalize
still further those of the already generalized gravitational
thermodynamics.
\end{abstract}



\maketitle

Twenty-one years ago, Morris, Thorne and Yurtsever
\cite{Morris:1988tu} presented the first classically consistent
solution for a wormhole that was stabilized by exotic matter and
became furthermore convertible into a time machine. That paper
meant a real breakthrough in that it inaugurated the history of
these space-time tunnels as real scientific objects, rather than
as science-fiction toys. In spite of that, wormholes remain still
belonging to the framework of technology fiction.
However, recent cosmological observations are compatible with the
dominance in the universal vacuum of a so called cosmic phantom
energy, which has been shown to play the same stabilizing role as
the conventional exotic matter with respect to wormholes
\cite{Sushkov:2005kj}.

It is well known that the thermodynamical description of
gravitational vacuum and black holes has provided these objects
with quite a more robust consistency, allowing moreover for a
deeper understanding of their space-time structure and properties.
We think that, since exotic matter can be viewed as the
time-reversed version of ordinary matter and hence one may well
think of wormholes as time-reversed black holes, if a similar
thermodynamical representation of wormholes would be feasible,
then the physical status of such tunneling would likewise greatly
improve, entitling us by the way to get a more comprehensible
account for the exotic matter which, on the other hand, would thus
become the largest energy source in the universe.

The aim of this paper is to construct a complete and physically
consistent wormhole thermodynamics, following the 2+2 formalism
developed by Hayward for dynamical black holes
\cite{Hayward:1993wb,Hayward:1997jp,Hayward:2004fz}, as that
formalism depends on local variables and can be in this way
applicable to space-times having no event horizon, such as that
for wormholes is. In fact, Hayward himself already realized that
this formalism may also be applied to wormholes
\cite{Hayward:1998pp}, as wormholes have an actual generalized
surface gravity \cite{Ida:1999an}, too. However, the important
contribution by Hayward to that subject lacks of a precise
definition of dynamic wormholes and of the formulation of the laws
of their thermodynamics, restricting himself to study the thermal
radiation just for the case of black holes. Also contemplated in
the present paper is the study of all these latter issues.

It is well known that in spherically symmetric space-times the
metric can generally be written as
\begin{equation}\label{uno}
{\rm d}s^2=2g_{+-}{\rm d}\xi^+{\rm d}\xi^-+r^2{\rm d}\Omega^2,
\end{equation}
where $\xi^\pm$ are the double-null coordinates and $r$ is the
areal radius. In terms of such an areal radius, the expansion
becomes
\begin{equation}\label{dos}
\Theta_{\pm}=\frac{2}{r}\partial_{\pm}r,
\end{equation}
with $\partial_{\pm}\equiv\partial/\partial\xi^{\pm}$ the two
preferred normal directions, which we shall consider to be
future-pointing.

In a spherically symmetric space-time one can introduce the Kodama
vector \cite{Kodama:1979vn}, which is defined by
\begin{equation}\label{tres}
k={\rm curl}_2r
\end{equation}
where the subscript $2$ means referring to the two-dimensional
space normal to the spheres of symmetry. An interesting property
of the Kodama vector, which eventually turns out to be similar to
that of the Killing vector, is \cite{Hayward:1997jp}
\begin{equation}\label{cuatro}
k\cdot\left(\nabla\wedge k^b\right)=\pm\kappa k^b
\end{equation}
on a trapping horizon, that is an hypersurface which can be
foliated by marginal spheres ($\Theta_+\Theta_-=0$) \cite{Hayward:1993wb}, at which
$k$ vanishes. In Eq. (\ref{cuatro}) $\kappa$ is the generalized
surface gravity
\begin{equation}\label{cinco}
\kappa=\frac{1}{2}{\rm div}_2{\rm grad}_2r,
\end{equation}
implying (choosing e.g. $\Theta_+=0$) that the outer
($\partial_-\Theta_+<0$), degenerated ($\partial_-\Theta_+=0$) and
inner ($\partial_-\Theta_+>0$) trapping horizons, respectively
have $\kappa>0$, $\kappa=0$ and $\kappa<0$.

In the case of a Morris-Thorne wormhole we would have a
bifurcating, i. e. $\Theta_{\pm}=0$, outer trapping horizon at the
wormhole throat $r_0$. Although this is a static case, no Killing
horizon is present. However one can then obtain a generalized
surface gravity given by
\begin{equation}\label{seis}
\kappa|_H=\frac{1-K'(r_0)}{4r_0}=-2\pi r_0(p+\rho)|_H,
\end{equation}
where $K(r)$ is the shape-function, $p$ the radial pressure,
$\rho$ the energy density and ``$|_H$'' means evaluation at the
horizon. The outward flaring condition implies $K'(r_0)<1$,
therefore the surface gravity is positive ($\kappa|_H>0$ because
$p+\rho<0$, equivalently), as it should be since we have an outer
horizon.

On the other hand, it is known that in a spherically symmetric
space-time a generalized first law of thermodynamics can be
written as \cite{Hayward:1997jp},
\begin{equation}\label{siete}
L_zE=\frac{\kappa L_zA}{8\pi}+\omega L_zV,
\end{equation}
where $A$ is the surface area, $L_z=z\cdot\nabla$ and
$z=z^+\partial_+ +z^-\partial_-$ is tangent to the trapping
horizon, $E$ is the Misner-Sharp energy given by
\begin{equation}\label{ocho}
E=\frac{1}{2}r\left(1-\partial^ar\partial_ar\right),
\end{equation}
and
\begin{equation}\label{nueve}
\omega=-\frac{1}{2}{\rm trace}_2T.
\end{equation}
Eq. (\ref{seis}) allows then to introduce an expression for the
geometric entropy, with the familiar dependence on the surface
area
\begin{equation}\label{diez}
S \propto A.
\end{equation}

In order to see how the area of a trapping horizon evolves, we can
express the surface area in terms of the 2-form area $\mu$ as
$A=\int_S\mu$, with $\mu=r^2\sin\theta{\rm d}\theta{\rm d}\varphi$
in the spherically symmetric case. It follows that the evolution
of a trapping horizon area is governed by the integral expression
\begin{equation}\label{once}
L_zA_H=\int_H\mu z^-\Theta_-,
\end{equation}
in which we have chosen $\Theta_+=0$ for the trapping horizon and
from the very definition of such a horizon $L_z\Theta_+=0$ should
keep taking on vanishing values along the entire horizon, so
implying
\begin{equation}\label{doce}
\frac{z^+}{z^-}=-\frac{\partial_-\Theta_+|_H}{\partial_+\Theta_+|_H}.
\end{equation}
The $++$ component of the Einstein equations, corresponding to
metric (\ref{uno}), the definition (\ref{dos}) and an energy
momentum tensor component $T_{++}$, can be written as \cite{Hayward:1994bu}
\begin{equation}\label{trece}
\partial_{+}\Theta_{+}=-\frac{1}{2} \Theta^2_{+}-\Theta_{+}\partial_{+}{\rm log}\left(-g_{+-}\right)
-8\pi T_{++},
\end{equation}
which, when evaluated at the trapping horizon, yields
\begin{equation}\label{catorce}
\partial_+\Theta_+|_H=-8\pi T_{++}|_H.
\end{equation}
It can be seen that by considering an energy-momentum tensor of type I according to the Hawking-Ellis classification \cite{HyE}, i. e. an energy
momentum tensor which takes a diagonal form in an orthonormal
basis, we have $T_{++}\propto p+\rho$ as expressed in our basis,
which actually is a very natural and consistent assumption\footnote{In general one would have $T_{++}\propto T_{00}+T_{11}-2T_{01}$, where the components of energy-momentum tensor on the r.h.s. are expressed with respect to an orthonormal basis. In our case, we consider an energy-momentum tensor of type I \cite{HyE}, not just because it represents all observer fields with non-zero rest mass and zero rest mass fields, except in special cases when it is type II \cite{HyE}, but also because if this would be not the case either $T_{++}=0$ (for types II and III) which at the end of the day would imply no horizon expansion, or we would be considering the case where the energy density vanishes (type IV)}.

Now, for a dynamical black hole which is defined by a future
($\Theta_-<0$) outer trapping horizon \cite{Hayward:1993wb} and
surrounded by ordinary matter with $p+\rho>0$, Eqs. (\ref{doce})
and (\ref{catorce}) will impose that the signs of non-vanishing
$z^+$ and $z^-$ ought to be different, so that the horizon is
space-like. Taking $z^+>0$, i. e. choosing $z$ with a positive
component along the future-pointing null direction of vanishing
expansion, it then follows from Eq.(\ref{once}) that $L_zA_H>0$
\cite{Hayward:2004fz}. A similar line of reasoning starting with
$p+\rho<0$ would finally lead to $L_zA_H<0$.

Babichev et al. \cite{Babichev:2004yx} used a test-fluid approach
to study the evolution of the horizon area of a Schwarzschild
black hole induced by the accretion of dark energy, and showed
similar results to those that we have just derived. If such
results are consistently assumed to be originated from a flow of
the surrounding matter into the hole, then one can regard both
methods to actually describe just the same single, physical
process. In what follows we shall consider that the above
coincidence is more than a mere analogy, so that it must also hold
in the case of wormholes. In fact, when applied to wormholes
\cite{GonzalezDiaz:2004vv}, the Babichev et al. procedure leads to
an increase (decrease) of the size of the wormhole throat in case
that $p+\rho<0$ ($p+\rho>0$). Such results can only be recovered
by using the $2+2$ formalism in terms of trapping horizons whereas
the outer trapping horizon of the wormhole is past ($\Theta_->0$
for $\Theta_+=0$, in which case $\xi^+$ would be ingoing and
$\xi^-$ outgoing). Or, in other words, in order to recover the same result following the $2+2$ formalism obtained by the mentioned procedure, i. e.
\begin{equation}\label{quince}
L_zA_H\geq0,
\end{equation}
for $p+\rho<0$ dominating the environment that surrounds a
wormhole, the wormhole must be characterized by a past outer trapping horizon\footnote{The same results can be obtained by using the
trapping horizon defined by $\Theta_-=0$, evolving according to
$L_z\Theta_-|_H=0$, and employing Einstein equations which,
evaluated at the horizon produces, $\partial_-\Theta_-|_H=-8\pi
T_{--}|_H$, since $T_{--}\propto \rho+p_r$, too.}. We would use in what follows this characterization.

Let us now consider the possible emission process associated with
the semi-classical effects, with the particle production rate
being given by the WKB approximation for the tunneling probability
$\Gamma$ along a classically forbidden trajectory,
$\Gamma\propto\exp\left[-2{\rm Im}\left(I\right)\right]$. If that
probability would take on a thermal form at the horizon,
$\Gamma\propto\exp\left(-\omega_\phi/T_H\right)$, then one could
easily obtain an expression for the temperature of that radiation.

Following then a parallel reasoning to that of Ref.
\cite{Hayward:2008jq} for black holes, we can express the metric
given by Eq.(\ref{uno}) in terms of the most convenient
generalized retarded Eddington-Finkelstein coordinates, owing to
the feature that a wormhole possesses a past outer trapping
horizon. That is
\begin{equation}\label{dieciseis}
{\rm d}s^2=-e^{2\Psi}C{\rm d}u^2-2e^\Psi{\rm d}u{\rm d}r+r^2{\rm
d}\Omega^2,
\end{equation}
where we have again considered $u=\xi^+$ related to the ingoing
direction, implying $\Theta_+|_H=0$, $\Theta_-|_H>0$ and
$\partial_-\Theta_+|_H<0$, ${\rm d}\xi^-=\partial_u\xi^-{\rm
d}u+\partial_r\xi^-{\rm d}r$, $e^\Psi=-g_{+-}\partial_r\xi^->0$
and $e^{2\Psi}C=-2g_{+-}\partial_u\xi^-$, with $C=1-2E/r$, $E$
defined by Eq. (\ref{tres}). $\Psi$ expresses all the gauge
freedom contained in the choice of the null coordinate $u$. It can
be noted that the use of retarded coordinates ensures that the
marginal surfaces, for which $C=0$, are past marginal surfaces. We want to emphasize that such a change of coordinates is general enough for our present purposes, assuming that the spacetime must possess a past outer trapping horizon; therefore it must be applicable to wormholes without any restriction about its traversability. On the other hand, a possible bad behaviour of the $r$ coordinate could be expected taking into account that the more natural and well behaved coordinate to describe the radial coordinate of a wormhole is $l$ (with $l$ such that $g_{ll}=1$ in orthogonal coordinates).

Let us also consider a massless scalar field in the eikonal
approximation, $\phi=\phi_0\exp\left(iI\right)$, with a slowly
varying amplitude and being governed by a rapidly varying action
given by \cite{Hayward:2008jq}
\begin{equation}\label{diecisiete}
I=\int \omega_\phi e^\Psi{\rm d}u-\int k_\phi{\rm d}r,
\end{equation}
in which $\omega_\phi$ and $k_\phi$ should be interpreted as the
angular frequency and wave number for the scalar field $\phi$,
respectively; that is $\partial_uI=\omega_\phi e^\Psi$ and
$\partial_rI=-k_\phi$. The field would then describe radially
outgoing radiation. Since the wave equation $\nabla^2\phi=0$
implies $g^{ab}\nabla_aI\nabla_bI=0$, one can finally obtain
\begin{equation}\label{dieciocho}
k_\phi^2C+2\omega_\phi k_\phi=0.
\end{equation}
This equation possesses two solutions: $k_\phi^{(1)}=0$, which
corresponds to the outgoing modes, and
$k_\phi^{(2)}=-2\omega_\phi/C$ for the ingoing modes.
$k_\phi^{(2)}$ produces a pole in the action (\ref{diecisiete}),
because $C|_H=0$ on the horizon. Noting that
$\kappa|_H=\partial_rC/2$, where we have taken $\Psi=0$, without
any loss of generality as $\kappa|_H$ should be gauge-invariant,
with expanding $C$, one obtains
$k_\phi\approx-\omega_\phi/\left[\kappa|_H(r-r_0)\right]$.
Therefore the action has an imaginary contribution which is
obtained by deforming the contour of integration in the upper $r$
half-plane, i.e. ${\rm
Im}\left(I\right)|_H=-\frac{\pi\omega_\phi}{\kappa|_H}$ and
$\Gamma$ has a thermal form for a temperature
\begin{equation}\label{diecinueve}
T=-\frac{\kappa|_H}{2\pi}.
\end{equation}
Of course, the radiation associated with temperature
(\ref{diecinueve}) has a semiclassical origin, being independent
of any possible classically allowed path passing through the
traversable wormhole throat. A rather key property of such a
temperature is its characteristic of being always negative, a
property stemming from the positiveness of the surface gravity on
the outer horizon, i. e. $\kappa|_H>0$. Some authors seem to be
rather uncomfortable with the concept of negative temperatures in
gravitational systems. While this attitude can well be
understandable for classical systems, it is not definitively so in
case of a quantum-mechanical system, where for sure one must not
be afraid of negative temperatures. In fact, different
experimentally checked devices which are inexorably interpreted in
terms of quantum-mechanically governed phenomena, have shown the
existence and properties of negative temperatures. On the other
hand, it is known that phantom energy has associated a negative
temperature \cite{GonzalezDiaz:2004eu}, so manifesting its quite
likely deep quantum nature. As we mentioned above, a phantom fluid
is one of the kinds of exotic materials which can be used as the
stuff to build up a traversable wormhole \cite{Sushkov:2005kj}.
Therefore, Eq.(\ref{diecinueve}) would imply that a wormhole
radiates ``particles'' with the same properties as its surrounding
matter, such as it already occurred with dynamical black holes in
relation to ordinary matter.

Eq. (\ref{diecinueve}) allows us to rewrite Eq.(\ref{siete}) in a
more familiar form which is given by
\begin{equation}\label{veinte}
L_zE=-TL_zS+\omega L_zV
\end{equation}\label{veintiuno}
on a trapping horizon, where
\begin{equation}\label{veintiuno}
S=\frac{A}{4}.
\end{equation}
The first term in the right hand side (r.h.s) of Eq.
(\ref{veinte}) can be interpreted as an energy-exchange term (in
analogy with the heat term of usual thermodynamics) and the second
one as a work-term. The negative sign in the energy-exchange term
would imply that the exotic matter which supports this space-time
gets energy from the space-time itself. Parallely, the term in the
r.h.s. of Eq.(\ref{veinte}) indicated that exotic matter carried
up a work in order to support the space-time. So we are already
prepared to formulate a {\it first law of wormhole thermodynamics}
in the following terms: {\it the change in the gravitational
energy of the wormhole is equal to the sum of the energy removed
from the wormhole and the work done in the wormhole.} Even more,
Eq. (\ref{veintiuno}) confirms relation (\ref{diez}), specifying
the involved proportionality constant which turns out to get the
familiar numerical value $1/4$, and, since $L_zA>0$ in an exotic
background, one has $L_zS>0$. That is to say, we have now a proper
{\it second law of wormhole thermodynamics expressible as: the
entropy of a wormhole, which is given in terms of the throat
surface area, can never decrease, when placed in its most natural
dominant-energy-condition violating environment.}

On the other hand, Eq.(\ref{cinco}) leads furthermore to a
formulation of the {\it third law of wormhole thermodynamics}, as
it implies that an outer horizon has always $\kappa>0$. If we
consider that no dynamical evolution is able to modify {\it the outer property of the horizon, then the generalized surface gravity would always remain being positive}. The above arguments can in fact be expressed by
stating: {\it if no dynamical process can change the outer character of the trapping horizon, then it is impossible to reach the absolute zero for
surface gravity by means of any dynamical process.}

It must be noted that if some dynamical
process could change the outer character of a trapping horizon in
such a way that it becomes an inner horizon, then the wormhole
would converts itself into a different physical object and,
therefore, the laws of wormhole thermodynamics would no longer be
valid.

It has been argued \cite{Hayward:1998pp} that by
replacing the background energy from exotic to ordinary, one also
changes the causal nature of an outer trapping horizon. It can
then be also considered that caused by such a process, or by a
subsequent one, a past outer trapping horizon (i. e. a dynamical
wormhole) should change into a future outer trapping horizon (i.e.
a dynamical black hole), and vice versa (check, for example, in
the case of the Schwarzschild solution).
The thermal radiation can be computed for a dynamical black
hole surrounded by ordinary matter by using the advanced
Eddington-Finkelstein coordinates, obtaining again Eq.
(\ref{diecinueve}) but with a plus sign, \cite{Hayward:2008jq}.
Therefore, if such a conversion would be at all possible, we
expected the temperature to also change from negative (wormhole)
to positive (black hole) in a way which is necessarily
discontinuous due to the holding of the third law, without passing
through the zero temperature.

Actually,
black holes and wormholes can be viewed to be the time-reversed
version of each other. It could seem that a white hole is the time-reversed version of a
black hole, but if one considers also the time inversion of the
surrounding material, then the surrounding ordinary matter would
become exotic matter and vice versa, changing the causal nature of
the horizon from space-like to time-like and vice versa. Such an idea about the relation between
wormholes and black holes under time-inversion is supported not
only by the fact that both are defined by an outer trapping
horizon, which is past (wormhole) and future (black holes), being
both bifurcating in the static case where no local dynamical
properties can be rigorously derived. Really, the three laws of
wormhole thermodynamics respectively become those of black hole
thermodynamics under time inversion. This can be easily checked
taking into account that the exotic matter is nothing but ordinary
matter moving backward in time, so implying a change in the sign
of temperature. Moreover, when one considers a time reversed first
law of wormholes thermodynamics, the energy-exchange term in Eq.
(\ref{veinte}) changes sign and the temperature becomes positive,
so that ordinary matter supplies energy to space-time. On the
other hand, we should also take into account that a dynamical
black hole emits radiation at a positive temperature. It is worth
noticing nevertheless that the other two wormhole thermodynamical
laws remain nevertheless unchanged under time inversion.

Finally, the growth of the wormhole area can be related to the
accretion of the surrounded exotic matter, but the radiation
process at the trapping horizon would produce a decrease of the
wormhole size, decreasing thereby the wormhole entropy, too. Like
in black holes thermodynamics such a violation of the second law
of thermodynamics is only apparent, because it is the total
entropy of the universe, composed by the wormhole, the surrounded
matter and the thermal radiation, what should increase. In fact,
one could perform an analysis similar to that followed in
\cite{Hawking:1976de} where a box filled with a black hole and
radiation was considered, in order to show that the entropy of the
whole system should always increase until the thermal equilibrium
is reached. The case of a box containing a wormhole just differs
from what is discussed in Ref. \cite{Hawking:1976de}, in that all
the contents in the box are at negative temperature, so that any
subsystem with lower temperature is hotter \cite{13}. Carefully
extending that analysis to the most general case of a box
simultaneously containing a black hole, a wormhole and given
amounts of ordinary and exotic radiations at different proportions
would lead to the conclusion that the interwoven final effect from
all possible involved thermal processes that can take place within
the box inexorably implies the holding of a most generalized
second law of thermodynamics for which the sum of all four
involved entropies always increases.

After completion of this paper we became aware of a paper by Hayward \cite{Hayward:2009yw} in which some part of the present work was also discussed following partly similar though somewhat divergent arguments. In work \cite{Hayward:2009yw} he studies the thermodynamic of two-types of dynamic wormholes, characterized by past or future outer trapping horizon. Although these two types are completely consistent mathematical solutions, we have concentrated on the present work in the first one since we consider that they are the only physical consistent wormholes solution. One of the reasons which support the previous claim has already been mentioned in this work and is based on the possible equivalence between the 2+2 formalism and the Babichev et al. method. On the other hand, a traversable wormhole must be supported by exotic matter and it is known that it can collapse by accretion of normal matter. That is precisely the problem of how to traverse a traversable wormhole finding the mouth open for the back-travel, or at least avoiding a possible death by a strangulated wormhole throat during the trip. If the physical wormhole could be characterized by a future outer trapping horizon, by Eqs. \ref{once}, \ref{doce} and \ref{catorce}, then it would increase (decrease) its size by accretion of ordinary (exotic) matter and, therefore, it would be not problem to traverse it, even more, it would increase its size when a traveler would pass through the wormhole, contrary to what it is expected for the basis of the wormhole physics.

\vspace*{0.5cm}

\acknowledgments

\noindent The authors are indebted to S. Robles-P\'{e}rez and A.
Rozas-Fern\'{a}ndez for useful discussions. P.~M.~M. gratefully
acknowledges the financial support provided by the I3P framework
of CSIC and the European Social Fund. This work was
supported by a Spanish MEC Research Project No. FIS2008-06332/FIS.


\begin{thebibliography}{99}
\bibitem{Morris:1988tu}
  M.~S.~Morris, K.~S.~Thorne and U.~Yurtsever,
  Phys.\ Rev.\ Lett.\  {\bf 61} (1988) 1446.
\bibitem{Sushkov:2005kj}
  S.~V.~Sushkov,
  Phys.\ Rev.\  D {\bf 71} (2005) 043520;
  F.~S.~N.~Lobo,
  Phys.\ Rev.\  D {\bf 71} (2005) 084011.

\bibitem{Hayward:1993wb}
  S.~A.~Hayward,
  Phys.\ Rev.\  D {\bf 49} (1994) 6467.

\bibitem{Hayward:1997jp}
  S.~A.~Hayward,
  Class.\ Quant.\ Grav.\  {\bf 15}, 3147 (1998).

\bibitem{Hayward:2004fz}
  S.~A.~Hayward,
  Phys.\ Rev.\  D {\bf 70}, 104027 (2004).

\bibitem{Hayward:1998pp}
  S.~A.~Hayward,
  Int.\ J.\ Mod.\ Phys.\  D {\bf 8}, 373 (1999).

\bibitem{Ida:1999an}
  D.~Ida and S.~A.~Hayward,
  Phys.\ Lett.\  A {\bf 260}, 175 (1999).

\bibitem{Kodama:1979vn}
  H.~Kodama,
  Prog.\ Theor.\ Phys.\  {\bf 63}, 1217 (1980).

\bibitem{Hayward:1994bu}
  S.~A.~Hayward,
  Phys.\ Rev.\  D {\bf 53} (1996) 1938

\bibitem{HyE}
  S.~W.~Hawking and G.~F.~R.~Ellis, {\it The large scale structure of space-time.},
  Cambridge University Press (1973).

\bibitem{Babichev:2004yx}
  E.~Babichev, V.~Dokuchaev and Yu.~Eroshenko,
  Phys.\ Rev.\ Lett.\  {\bf 93} (2004) 021102
  E.~Babichev, V.~Dokuchaev and Y.~Eroshenko,
  J.\ Exp.\ Theor.\ Phys.\  {\bf 100} (2005) 528
  [Zh.\ Eksp.\ Teor.\ Fiz.\  {\bf 127} (2005) 597]

\bibitem{GonzalezDiaz:2004vv}
  P.~F.~Gonz\'alez-D\'{\i}az,
  Phys.\ Rev.\ Lett.\  {\bf 93} (2004) 071301
  P.~F.~Gonz\'alez-D\'{\i}az and P.~Mart\'{\i}n-Moruno,
{\it Proceedings of the eleventhMarcel Grossmann Meeting on General Relativity}, Editors: H.~Kleinert, R.~T.~Jantzen and R.~Ruffini, World Scientific, New Jersey (2008)

\bibitem{Hayward:2008jq}
  S.~A.~Hayward, R.~Di Criscienzo, L.~Vanzo, M.~Nadalini and S.~Zerbini,
  arXiv:0806.0014 [gr-qc].

\bibitem{GonzalezDiaz:2004eu}
  P.~F.~Gonzalez-Diaz and C.~L.~Siguenza,
  Nucl.\ Phys.\  B {\bf 697}, 363 (2004)
  E.~N.~Saridakis, P.~F.~Gonzalez-Diaz and C.~L.~Siguenza,
  arXiv:0901.1213 [astro-ph].

\bibitem{Hawking:1976de}
  S.~W.~Hawking,
  Phys.\ Rev.\  D {\bf 13} (1976) 191.

\bibitem{13}
  A.~Alonso and P.~Martin-Moruno, Work in Preparation.

\bibitem{Hayward:2009yw}
  S.~A.~Hayward,
  arXiv:0903.5438 [gr-qc].



\end{thebibliography}
\end{document}